\documentclass[journal,comsoc,twocolumn]{IEEEtran}

\ifCLASSINFOpdf
  
\else
  
\fi

\usepackage{ifpdf}
\usepackage{cite}
\usepackage{amsmath}
\usepackage{algorithmic}
\usepackage{algorithm}
\usepackage{multicol}
\usepackage{array}

\usepackage{amssymb,amsfonts}
\usepackage{graphicx}
\usepackage{textcomp}
\usepackage{xcolor}
\usepackage{subcaption}
\usepackage[dvips]{epsfig}
\usepackage{kotex}
\begin{document}

\title{Robust and Low Complexity Beam Tracking with Monopulse Signal for UAV Communications}

\author{Ha-Lim Song,~\IEEEmembership{Student Member,~IEEE,}~and Young-Chai Ko,~\IEEEmembership{Senior Member,~IEEE}
\thanks{
$^{\star}$ H.-L. Song and Y.-C. Ko are with Korea University, Seoul, Korea. Y.-C. Ko is a corresponding author (Email: \{hhhalims,koyc\}@korea.ac.kr).
}}

\maketitle

\begin{abstract}
UAV communications based on an antenna array entail a beam tracking technology for reliable link acquisition. Unlike conventional cellular communication, beam tracking in UAV communication addresses new issues such as mobility and abrupt channel disconnection from UAV's perturbation. To deal with these issues, we propose a beam tracking scheme based on extended Kalman filter (EKF) using a monopulse signal, which can provide (1) higher robustness by offering a reliable link in the estimated spatial direction and (2) lower complexity compared with the existing conventional beam tracking schemes. We point out the limitations of using a beamformed signal as a measurement model for a Kalman filter (KF) based scheme and instead utilize the monopulse signal as a more plausible model. For the performance evaluation, we derive the upper bound of the mean square error for spatial angle estimation of UAV and confirm that the proposed scheme is stable with a certain bounded error. We also show from various simulations that the proposed scheme can efficiently track UAV and detect beam disconnection every 
time frame using a beamformed signal.
\end{abstract}


\IEEEpeerreviewmaketitle

\section{Introduction}
With the increasing demands for applications of unmanned aerial vehicle (UAV) communications as an element technology of 5G or B5G, the critical issue is to have a reliable link between the ground node and a UAV or UAV-to-UAV \cite{xiao2016enabling,zhao2018beam,geraci2018understanding}. The UAV communication assisted link can be efficiently established by aligning the beam toward UAV because the line of sight (LOS) is mainly considered in the aerial network \cite{cuvelier2018mmwave}. Note that the beam alignment between a transmitter and a receiver and an accurate direction parameter estimation are two prerequisites for a beam tracking \cite{zhao2018channel,razi2019optimal,zhou2018beam}.

One of the main beam tracking schemes is based on Kalman filter (KF) \cite{zhang2016tracking,liu2018channel,va2016beam,jayaprakasam2017robust,kim2019robust,larew2019adaptive,zhao2018channel}. It is widely known that two KF based algorithms, extended Kalman filter (EKF) \cite{ljung1979asymptotic} and unscented Kalman filter (UKF) \cite{wan2000unscented}, can be applied to the nonlinear model. Most studies of KF based beam tracking have established the beamformed signal as a measurement model \cite{zhang2016tracking,liu2018channel,va2016beam,jayaprakasam2017robust,larew2019adaptive}. They focus on the received beamformed signal which is described as adaptive beamforming \cite{liu2018channel,va2016beam,jayaprakasam2017robust,kim2019robust} or codebook based beamforming  \cite{zhang2016tracking,larew2019adaptive}. However, 
it is not suitable for a KF based scheme, which provides an optimal solution in a linear system, because the nonlinearity of the beamformed signal is strong. The considered measurements such as real and imaginary terms of the received signal follow $\frac{\sin{Nx}}{\sin{x}}$ function. Even if we only deal with the main lobe of the beam, it is a concave function 
of spatial angle $x$, which lies within the main lobe beamwidth. Therefore, it should be noted that the performance of EKF and UKF based beam tracking schemes with the measured beamformed signal may be limited since such schemes require linear approximation.

More importantly, their measurement models are affected by the parameters of the beamforming weight, $\hat{\phi}$, as well as the angular parameters of the channel, $\phi$. When it comes to adaptive beamforming deriving the beam toward a particular direction as $\hat{\phi}$, the KF based scheme can be a viable solution only if the derived beam targets the main lobe of beam pattern corresponding to the actual channel. Thus, determining the beam  direction is not practical toward the main lobe of the beam pattern without the actual channel state information. \cite{va2016beam,jayaprakasam2017robust,liu2018channel,kim2019robust}. In \cite{liu2018channel,jayaprakasam2017robust}, authors tackle additional optimization methods to compensate for the performance loss of EKF, but that do not 
resolve the underlying problem of designing the beamforming weight parameter, $\hat{\phi}$. The authors of \cite{kim2019robust} utilize the auxiliary beam pair method \cite{zhu2017auxiliary} based on the amplitude comparison monopulse method \cite{sherman2011monopulse}, but correct beam selection is required to achieve high estimation performance. Note that the auxiliary beam pair method uses the measurement model of the ratio metric, not the beamformed signal itself, so that the nonlinearity can be mitigated.
The codebook based beamforming scheme \cite{zhang2016tracking,larew2019adaptive} shows a more stable performance than the former \cite{va2016beam,jayaprakasam2017robust,liu2018channel}, but steering the entire codebook every time index is not practical. Moreover, the dimension of the measurement model becomes very large, which requires $\mathcal{O}\left(n^3\right)$ complexity to calculate the innovation covariance matrix or Kalman gain, where $n$ is related to the size of the codebook. The works in \cite{larew2019adaptive} employ the UKF and conduct extra optimization to reduce the size of codebook, but it also raises overhead to find a sub-optimal beamforming matrix. Besides them, there are sensor based beam tracking researches \cite{zhao2018channel,liu2019learning}, and the authors in \cite{zhao2018channel,liu2019learning} apply UKF on estimating the position information of UAV with sensor measurements such as global positioning system (GPS) and flight controller.

This paper focuses on three main contributions to UAV communications.
1) We propose a beam tracking scheme based on EKF using the monopulse signal, which is a more plausible measurement model of KF for lower complexity compared to beamformed signal based KF algorithms.
2) It is observed that the proposed scheme can efficiently track the spatial angle of UAV and detect error and beam misalignment over time frame. 3) To evaluate the performance, we derive the upper bound of mean square error (MSE) and verify that the proposed scheme is stable under the bounded estimation error over time while UAV keeps flying. Simulation results show that the proposed scheme outperforms the conventional schemes \cite{zhang2016tracking,larew2019adaptive,kim2019robust} under planar array and LOS channel. 

The rest of paper is organized as follows. In Sec. \ref{sys model}, we present the system and channel models under consideration. In Sec. \ref{body_algo}, we propose the robust EKF based beam tracking algorithm with monopulse signal measurement, and in Sec. \ref{body_dt}, we present the error detection method for the stable beam connection. In Sec. \ref{body_analy}, we analyze the performance in terms of bounded mean square error for tracking, and then we present some selected results in Sec. \ref{simulation}. Sec. \ref{conclusion} summarizes the proposed scheme.

\section{System and Channel Models} \label{sys model}
\begin{figure}[htbp]
    \centering
    \includegraphics[width=0.45\textwidth]{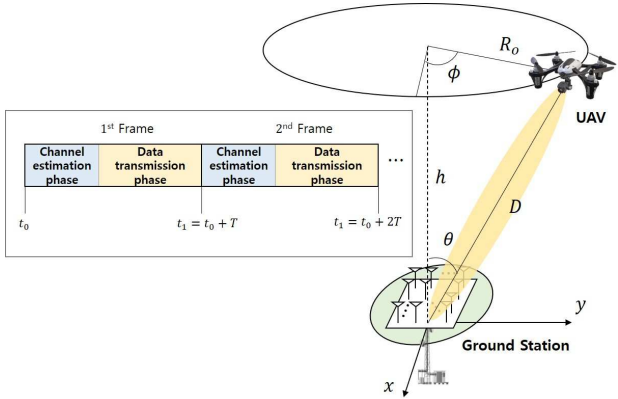}
    \caption{System model of UAV communication with the two phases of frame, where the GS is equipped with a $N_x\times N_y$-UPA, while the UAV has a single antenna.}
    \label{fig:sys}
\end{figure}

We consider a beam tracking system as depicted in Fig. \ref{fig:sys}. At each frame, ground station (GS) estimates the channel parameter based on the monopulse signal, and then GS transmits/receives information-bearing signal data to/from the target UAV. Here, we assume that the channel is invariant during $T$ period of a frame. The frame comprises two phases, such as 1) channel estimation phase and 2) data transmission phase. In the channel estimation phase, we estimate the spatial angle by using the pilot signal and align the beam to the estimated parameter. We transmit/receive the information data to/from the target and detect beam alignment with beamformed signal in the transmission phase. If the beam misalignment over a certain error threshold is continuously detected, we implement a mechanical alignment and initialize the process.

We consider a GS with $N=N_x \times N_y$ uniform planar array (UPA) and target UAV, which is tracked by GS, has a single antenna. Note that the proposed scheme can be easily extended to multiple antenna structure of UAV by considering the angle of departure of transmitted signals on the array of UAV. Moreover, the LOS channel link is mainly assumed in the high altitude air network that the majority of the signals are transmitted by the LOS path with high probability \cite{liu2018channel,zhao2018channel}. Then, we have a time-varying channel at the $k$th frame between GS and UAV with distance $D_k$, which can be represented as 
\begin{equation} \label{eq.h}
    \mathbf{H}_k = \frac{\rho_k\alpha_k}{D_k^{\beta_k}}\mathbf{a}_x(u_k)\mathbf{a}_y(v_k)^H,
\end{equation}
where $(\cdot)_k$ denotes the corresponding value of the $k$th frame. In (\ref{eq.h}), $\rho_k$ denotes the path-loss gain embracing the antenna gain and transmitted power, and $\alpha_k$ and $\beta_k$ are the channel gain and path-loss exponent, respectively. We also denote $u_k$ and $v_k$ for spatial direction parameter of UAV's transmitted signal for $x$-axis and $y$-axis, respectively.
The array response vector for $x$-axis and $y$-axis, $\mathbf{a}_x$ and $\mathbf{a}_y$, in (\ref{eq.h}) can be expressed as
\begin{equation} \label{eq.arr_x}
 \mathbf{a}_x(u_k)= \begin{bmatrix} 1\! & \!e^{-ju_k}\!& \!\cdots\! &\!e^{-j(N_x-1)u_k}\end{bmatrix}^T,
\end{equation}
\begin{equation} \label{eq.arr_y}
 \mathbf{a}_y(v_k)= \begin{bmatrix} 1\! & \!e^{-jv_k}\!& \!\cdots\! &\!e^{-j(N_y-1)v_k}\end{bmatrix}^T,
\end{equation}
respectively, where $u_k$ and $v_k$ denote the spatial angle as $\frac{2\pi d}{\lambda}\cos{\phi_k}\sin{\theta_k}$ and $\frac{2\pi d}{\lambda}\sin{\phi_k}\sin{\theta_k}$ with $d$ of antenna spacing, and $\lambda$ of the wavelength of the incident signal. The azimuth and elevation angles for the incoming signal of UAV are denoted as $\phi_k$ and $\theta_k$, respectively as in Fig. \ref{fig:sys}. Considering the UAV flying the sky at the constant altitude for particular missions \cite{yang2019beam}, we can set up the state model for spatial angle as
\begin{equation} \label{eq:state}
    \begin{aligned}
    \mathbf{x}_{k+1} =& \begin{bmatrix}u_{k+1}\\v_{k+1}\end{bmatrix}=\begin{bmatrix}\cos{\psi}& -\sin{\psi}\\ \sin{\psi} &\cos{\psi}\end{bmatrix}\begin{bmatrix}u_{k}\\v_{k}\end{bmatrix}+ \begin{bmatrix}\omega_{u,k}\\\omega_{v,k}\end{bmatrix}, \\
    =&\mathbf{F}\mathbf{x}_{k}+\mathbf{n}_{p,k},
    \end{aligned}
\end{equation}
where $\omega_{u,k}$ and $\omega_{v,k}$ are the process noise with variances $\sigma_u^2$ and $\sigma_v^2$, respectively, and $\mathbf{n}_{p,k}$ denotes the corresponding process noise vector at the $k$th frame.
$\psi$ is the angle of rotation of the UAV every time frame. We first propose the state evolution model for a practical UAV's movement and the obtained state can be directly employed to designing the beamforming weight. 
Moreover, the channel gain $\alpha_k$ is given by the first-order Gaussian-Markov model \cite{va2016beam}
\begin{equation}
    \alpha_{k+1}=\rho\alpha_k+\epsilon_{k},
\end{equation}
where $\rho$ is the correlation coefficient and $\epsilon_{k}$ follows the normal distribution as $\mathcal{CN}(0,(1-\rho^2/2))$.
Then we can write the received signal in matrix form as
\begin{equation} \label{eq:mes}
    \mathbf{Y}_k = {\mathbf{H}}_k s_k + \mathbf{N}_k,
\end{equation}
where $s_k$ denotes a single transmitted pilot signal at a certain time in the $k$th frame, and $\mathbf{N}_k$ is the complex white Gaussian noise matrix. The beamformed signal in the transmission phase of the $k$th frame can be represented by
\begin{equation} \label{eq:mes}
    r_k = \mathbf{w}^H(\hat{\mathbf{x}}_k) {\mathbf{h}}_k s_d + \mathbf{w}^H(\hat{\mathbf{x}}_k)\mathbf{n}_k,
\end{equation}
where $s_d$ denotes the data-bearing signal, and
$\hat{\mathbf{x}}_k$ denotes the estimated angle of the target. 
The corresponding beamforming vector for the estimated angle can be denoted as $\mathbf{w}(\hat{\mathbf{x}}_k)$. In (\ref{eq:mes}), $\mathbf{h}_k$ and $\mathbf{n}_k$ denote the channel and noise in column vector form, respectively.
 The beamforming weight vector $\mathbf{w}(\hat{\mathbf{x}}_k)$ generates the beam for the desired direction and can be written as
\begin{equation}
    \mathbf{w}(\hat{\mathbf{x}}_k)=\mathrm{vec}\left(\mathbf{w}_x(\hat{u}_k) \mathbf{w}_y(\hat{v}_k)^H\right),
\end{equation}
where $\mathrm{vec}(\cdot)$ represents transformation from matrix to vector and the beamforming weight vector for $x$-axis and $y$-axis can be defined, respectively, as  
\begin{equation} \label{eq:w_x}
\!\!\!\!\!\!\mathbf{w}_x(\hat{u}_k)= \frac{1}{\sqrt{N_x}} \begin{bmatrix} 1\! & \!e^{-j\hat{u}_k}\!& \!\cdots\! &\!e^{-j(N_X-1)\hat{u}_k}\end{bmatrix}^T,
\end{equation}
\begin{equation} \label{eq:w_y}
\mathbf{w}_y(\hat{v}_k)= \frac{1}{\sqrt{N_y}} \begin{bmatrix} 1\! & \!e^{-j\hat{v}_k}\!& \!\cdots\! &\!e^{-j(N_y-1)\hat{v}_k}\end{bmatrix}^T.
\end{equation}

\section{Proposed Beam Tracking based on EKF with Monopulse Signal} \label{body_algo}
The KF generally assumes the Gaussian distribution for both linear state transition and measurement models based on Markov property, so the distribution of transitional state is conditionally defined by the previous state as
 $p(\mathbf{x}_{k}|\mathbf{x}_{k-1})=\mathcal{N}(\mathbf{x}_{k}|\mathbf{F}\mathbf{x}_{k-1},\mathbf{Q}_{p,k})$, and the measurement likelihood distribution can be derived as
$p(\mathbf{y}_{k}|\mathbf{x}_{k})=\mathcal{N}(\mathbf{y}_{k}|\mathbf{G}\mathbf{x}_{k},\mathbf{Q}_{n,k})$ \cite{haykin2004kalman}. $\mathbf{G}$ is a measurement model, and $\mathbf{Q}_{p,k}$ and $\mathbf{Q}_{n,k}$ are process and measurement noise matrices at the $k$th frame, respectively. We herein denote $\mathcal{N}(\mu,\sigma^2)$ as Gaussian distribution with mean, $\mu$ and variance, $\sigma^2$.
Thus, KF in a linear system has an optimal recursive solution as $\hat{\mathbf{x}}_k=\it{E}[\mathbf{x}_k|\mathbf{y}_{1:k}]$ given the observations from a minimum mean square error (MMSE) point of view.
The existing researches mainly determine the received beamformed signal as the measurement model in KF-based beamforming system  \cite{zhang2016tracking,larew2019adaptive,liu2018channel,va2016beam,jayaprakasam2017robust}. However, this beamforming signal model is highly nonlinear as in the form of $\frac{\sin{(Nx)}}{\sin{x}}$, resulting in performance degradation. As an alternative, we set up the measurement model with a nonlinear but more plausible monopulse signal. 

The monopulse signal is defined as the ratio of sum and difference of the received signals in two adjacent antennas \cite{song2019aoa}, which is derived from complex comparison monopulse method \cite{sherman2011monopulse}. We point out that the considered method is different from auxiliary beam pair method for an antenna array, which is based on the amplitude comparison monopulse method used in the radar systems \cite{sherman2011monopulse,zhu2017auxiliary,kim2019robust}. The proposed scheme can enable fast angle estimation on received signals, not on beamformed signals, drastically reducing the beam codebook search complexity compared to the conventional beam training schemes. The key differences with conventional beam tracking schemes are explained in detail in Sec. \ref{body_algo}. A.

Given the simplest case, when the received signals in the uniform linear array are defined as $\begin{bmatrix}1 & e^{-j u} & e^{-j2u} &\dots &e^{-j(N-1)u}\end{bmatrix}^T$, the monopulse signal $R$ can be expressed as 
\begin{equation} \label{eq:R}
    R=\frac{1-e^{-j u}}{1+e^{-j u}}= j\frac{\sin{\frac{u}{2}}}{\cos{\frac{u}{2}}}= j\tan{\frac{u}{2}}.
\end{equation}
Then, we can obtain the phase difference $u$ from $\arctan({\Im\{R\}})$, where $\Im$ denotes imaginary term of the corresponding complex value.
In the proposed system with a planar array, we extend (\ref{eq:R}) to the monopulse signals on the $x$-axis and $y$-axis by denoting $R_x$ and $R_y$, which can be written, respectively, as
\begin{equation} \label{Rx_def}
\begin{aligned}
    R_{x,k}&=\frac{1}{N_y(N_x-1)}\sum_{m=1}^{N_y}\sum_{n=1}^{N_x-1}\frac{\bar{\mathbf{Y}}_k(n,m)-\bar{\mathbf{Y}}_k(n+1,m)}{\bar{\mathbf{Y}}_k(n,m)+\bar{\mathbf{Y}}_k(n+1,m)},\\
    & \simeq j\tan{\left(\frac{u}{2}\right)},
\end{aligned}
\end{equation}
\begin{equation} \label{Ry_def}
\begin{aligned}
    R_{y,k}&=\frac{1}{N_x(N_y-1)}\sum_{n=1}^{N_x}\sum_{m=1}^{N_y-1}\frac{\bar{\mathbf{Y}}_k(n,m)-\bar{\mathbf{Y}}_k(n,m+1)}{\bar{\mathbf{Y}}_k(n,m)+\bar{\mathbf{Y}}_k(n,m+1)}, \\
    & \simeq j\tan{\left(\frac{v}{2}\right)},
\end{aligned}
\end{equation}
where $\bar{\mathbf{Y}}_k(n,m)$ denotes a normalized received signal of ($n$, $m$)-th antenna element of the planar array in the $k$th frame, whose magnitude and path loss effect are assumed to be mitigated by automatic gain control. 
Given that $\Im\{R_x\}=\tan{(u/2)}$ and $\Im\{R_y\}=\tan{(v/2)}$, the measurement model with monopulse signal can be set up as
\begin{equation} \label{eq:meas_r}
    \mathbf{r}_k = \begin{bmatrix}\tan{\left(\frac{u_k}{2}\right)}\\\tan{\left(\frac{v_k}{2}\right)}\end{bmatrix} + \begin{bmatrix}n_{u,k}\\n_{v,k}\end{bmatrix}=g\left(\mathbf{x}_k\right)+ \mathbf{n}_{m,k},
\end{equation}
where $n_{u,k}$ and $n_{v,k}$ are measurement noise with variances $\sigma_n^2$.
Note that EKF is the extended version of KF to address the nonlinear function of state and measurement models by linearizing the nonlinear function by the first-order approximation \cite{jazwinski2007stochastic}. Here, we adopt EKF to linearize the nonlinear measurement model based on monopulse signal. 
\begin{algorithm} [t]
\caption{EKF based Robust Beam Tracking Algorithm} \label{algo1}
\begin{algorithmic} [1]
\STATE Initial estimation 
\FOR{$1\le k\le T_{N}$} 
\STATE \textbf{Channel estimation phase}
\STATE Predict the state, $\hat{\mathbf{x}}_k^-$ and the covariance matrix of the state, $\mathbf{P}_k^-$ in (15), (16)
\STATE Measure the monopulse signal, $\mathbf{r}_k$
\STATE Obtain the innovation, $\tilde{\mathbf{r}}_k$ and innovation covariance matrix $\mathbf{S}_k$ in (17)
\STATE Calculate the Kalman gain $\mathbf{K}_k$ in (18)
\STATE Update and correct the predicted state, $\hat{\mathbf{x}}_k$ and covariance matrix, $\mathbf{P}_k$ in (19)
\STATE \textbf{Data transmission phase}
\STATE Transmit/Receive information-bearing signal to/from UAV
\STATE Beam misalignment detection using beamformed signal and obtain the estimated error
\IF{$\hat{\lVert \boldsymbol{\xi} \rVert}_k>p_{th}$}
\STATE Execute mechanical alignment 
\STATE Initial estimation
\ENDIF
\STATE $k=k+1$
\ENDFOR
\end{algorithmic}
\end{algorithm}

\begin{figure}[htbp]
    \centering
    \includegraphics[width=0.4\textwidth]{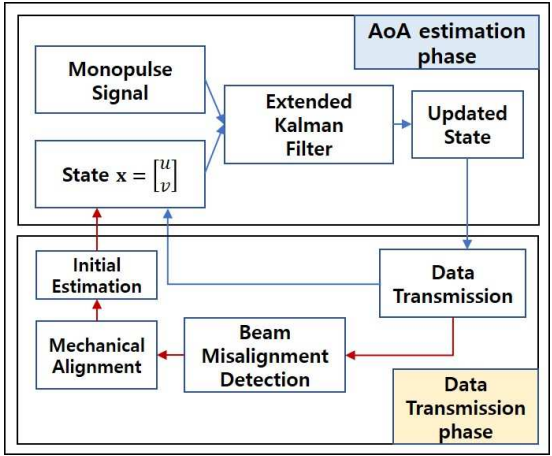}
    \caption{Robust EKF based beam tracking protocol}
    \label{fig:algo}
\end{figure}
For the first step to EKF, the predicted state $\hat{\mathbf{x}}_k^-$ can be calculated by using the previous estimated state $\hat{\mathbf{x}}_{k-1}$ as 
\begin{equation} \label{eq:pred}
    \hat{\mathbf{x}}_k^-=\mathbf{F}\hat{\mathbf{x}}_{k-1}.
\end{equation}
The corresponding predicted error covariance is defined as 
\begin{equation}
    \mathbf{P}_k^-=\mathbb{E}[(\mathbf{x}_k-\hat{\mathbf{x}}_k^-)(\mathbf{x}_k-\hat{\mathbf{x}}_k^-)^T] = \mathbf{F}\mathbf{P}_{k-1} \mathbf{F}^T+\mathbf{Q}_{p,k},
\end{equation}
where $\mathbf{P}_{k-1}$ is the previous state error covariance.
In the next step for update and correction, we need to obtain the innovation as 
\begin{equation}
\begin{aligned} \label{eq:G}
    \tilde{\mathbf{r}}_k=\mathbf{r}_k-g(\hat{\mathbf{x}}_k^-) &= g(\mathbf{x}_k)+\mathbf{n}_{m,k}-g(\hat{\mathbf{x}}_k^-), \\
    &\simeq \mathbf{G}\mathbf{x}_k+\mathbf{n}_{m,k}-\mathbf{G}\hat{\mathbf{x}}_k^-,  
\end{aligned}
\end{equation}
where $g(\hat{\mathbf{x}}_k^-)$ denotes predicted measurement value for the predicted state, $\mathbf{G}$ is the Jacobian matrix as $\left.\frac{\partial g(\mathbf{x})}{\partial \mathbf{x}}\right\vert_{\hat{\mathbf{x}}_{k}^{-}}\simeq0.5\mathbf{I}_2$ where $\mathbf{I}_2$ is the $2\times2$ indentity matrix, and $\mathbf{n}_k$ is measurement noise vector in (\ref{eq:meas_r}).
Then, the innovation covariance of $\tilde{\mathbf{r}}_k$ can be expressed as $\mathbf{S}_k=\mathbf{G}\mathbf{P}_k^-\mathbf{G}^T+\mathbf{Q}_{n,k}$. Here, since the measurement dimension of this scheme is $2$ regardless of the number of antenna elements, the proposed scheme can significantly reduce the complexity related to the size of antenna array. 
The Kalman gain then can be defined as 
\begin{equation} \label{eq:kalman_g}
    \mathbf{K}_k=\mathbf{P}_k^-\mathbf{G}^T\mathbf{S}_k^{-1},
\end{equation}
which can minimize the MSE of estimation.
Finally, the updated state and error covariance are obtained by 
\begin{equation} \label{eq:update}
    \begin{aligned}
    \hat{\mathbf{x}}_k=&\hat{\mathbf{x}}_k^-+\mathbf{K}_k\tilde{\mathbf{r}}_k, \\
    \mathbf{P}_{k}=&\mathbf{P}_k^--\mathbf{K}_k\mathbf{S}_k\mathbf{K}_k^T.
    \end{aligned}
\end{equation}
EKF repeats the prediction and the update-correction steps in every estimation phase, as illustrated in Fig. \ref{fig:algo}. As such, the system can establish a reliable communication link with 3D beamforming gain for the updated state, which will be shown in the following sections.

\subsection{The plausibility of the monopulse signal in EKF based beam tracking applications}
\begin{figure}
    \centering
    \subcaptionbox{Measurement model of the monopulse signal}{\includegraphics[width=0.5\textwidth]{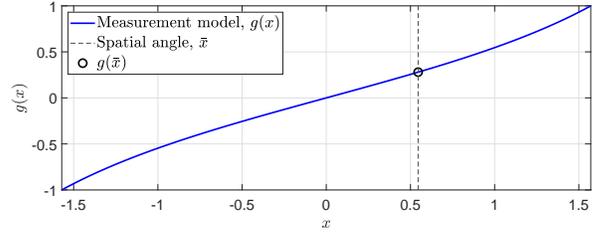}}
    \centering
    \subcaptionbox{Measurement model of the codebook based beamforming \\ signal}{\includegraphics[width=0.5\textwidth]{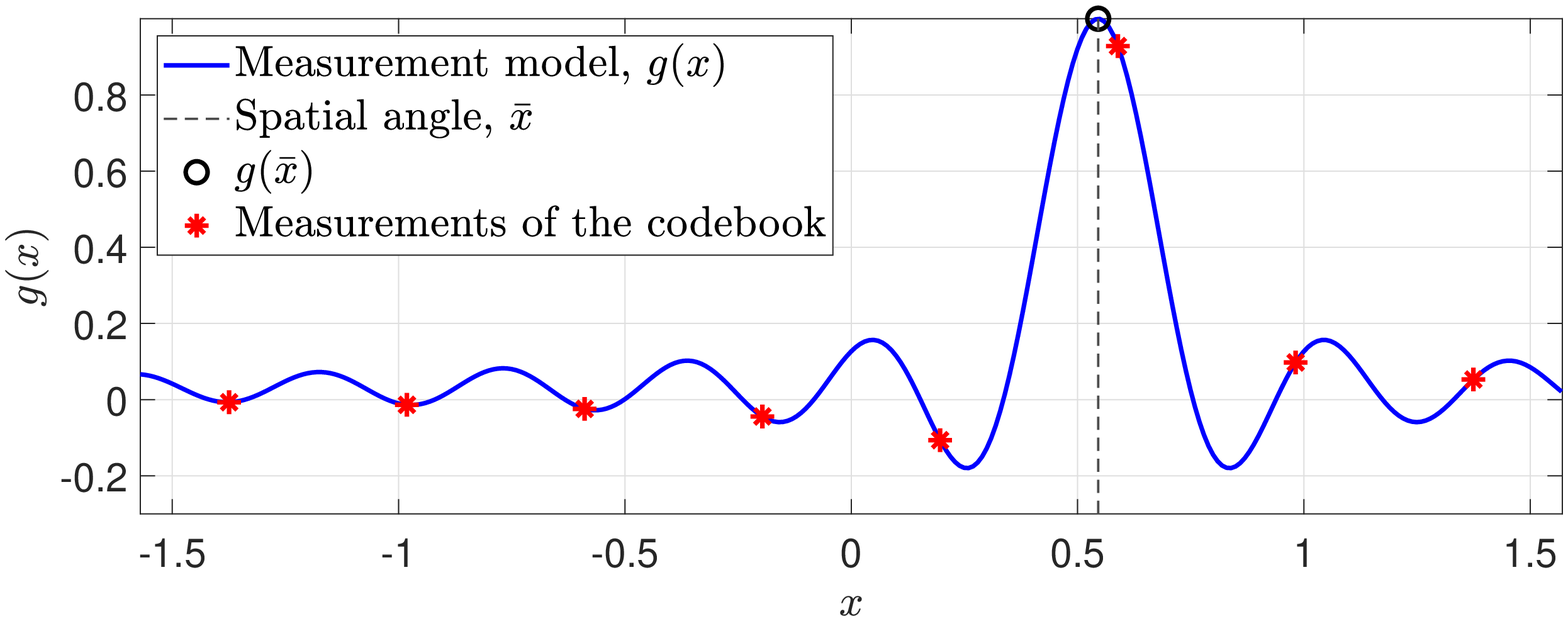}}
    \centering
    \subcaptionbox{Measurement model of the auxiliary beam pair signal}{\includegraphics[width=0.5\textwidth]{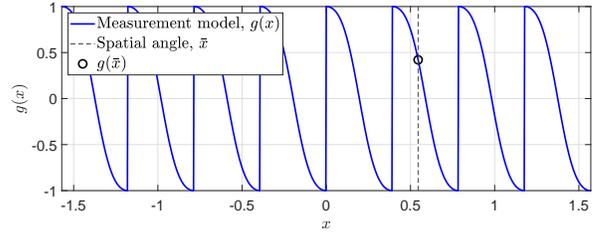}}
\caption{Measurement model for proposed scheme versus other schemes with spatial angle $\bar{x}=0.5455$, and codebook size, $K=8$, and the number of antennas, $N=8$.}
\label{fig:meas_model}
\end{figure}
Let us first explain the plausibility of the monopulse signal on the EKF based beam tracking applications by comparing the proposed scheme with the conventional schemes such as codebook based beamforming \cite{zhang2016tracking,larew2019adaptive} and auxiliary beam pair \cite{zhu2017auxiliary,kim2019robust}. We consider an $8$-uniform linear array (ULA) to represent the measurement model in a 1-dimensional plot for simplicity and $8$-sized beam codebook. The spatial angle of the channel is $\bar{x}=0.5455$.
\subsubsection{Measurement model of EKF}
Fig. \ref{fig:meas_model} shows the measurement models of Kalman based beam tracking schemes under consideration. 
The Kalman gain of EKF can be calculated in (\ref{eq:G}) and (\ref{eq:kalman_g}) by linearizing the nonlinear measurement function. This can generate performance degradation. Firstly, the complex comparison monopulse method for an array can derive the measurement model as in (\ref{Rx_def}). 
In Fig. \ref{fig:meas_model} (a), we show that the proposed scheme can apply all angles from $-\pi/2$ to $\pi/2$ to the corresponding function. Therefore, it is less susceptible to errors resulting from the normal vector of the array towards the target\footnote{The estimation performance can be degraded when the value of spatial angle deviates far from the normal vector of the array, so we address the issue in the following Sec. \ref{body_dt}.} than the amplitude comparison monopulse method. Moreover, the linearization loss is relatively small as $\tan(x)\simeq \frac{1}{2}x$. In addition, Fig. \ref{fig:meas_model} (b) shows the measurement model of the codebook based beamforming scheme. This scheme should calculate the Jacobian matrix $\mathbf{G}$ at eight beamformed signals over codebook angles as in (\ref{eq:G}). However, the codebook based beamforming scheme would be hard to correct the predicted state in the update step in (\ref{eq:update}), considering the strong non-linearity of the corresponding function. 
In Fig. \ref{fig:meas_model} (c), we can see the measurement model of the auxiliary beam pair scheme, which is a ratio metric based on the amplitude comparison monopulse method.
The particular pattern is repeated periodically, where the ratio metric function can be obtained based on the individual eight center beams. It should select the one beam among the eight center beams, calculating the Jacobian matrix. Thus, wrong beam selection can make the predicted state not to be appropriately corrected. 
\subsubsection{Estimation and calculation overheads}
The complex comparison monopulse method for an array can estimate the spatial angle at once by using the received signal without beamforming overhead. The size of the measurement model is $1$ in the ULA case, which is extended to $2$ in the UPA case as in (\ref{eq:meas_r}). The small-sized measurement model can reduce the computational burden regarding the inversion of the innovation covariance matrix in (\ref{eq:kalman_g}). Moreover, the calculation process required to obtain the monopulse signal in (\ref{Rx_def}) is comprised of arithmetic functions, which can be negligible in terms of calculation complexity. 
In addition, the size of the measurement of the codebook based beamforming scheme depends on the size of the codebook, $K=8$, which is related to the number of antennas. The real and imaginary terms of the received beamformed signals are employed in EKF, so the total measurement size becomes $m=2\cdot K$ in the ULA case. The high-resolution codebook can increase the calculation complexity drastically as $\mathcal{O}(m^3)$ in (\ref{eq:kalman_g}). Moreover, the codebook based beamforming requires the beamforming overhead for the corresponding codebook.
Finally, the auxiliary beam pair scheme has the same sized measurement model as the proposed scheme, but it also requires beamforming overheads for the corresponding codebook and adjacent pair beams.

\section{Beam Misalignment Detection} \label{body_dt}
We consider two procedures for the beam alignment such as 1) coarse alignment and 2) fine alignment for UAV communications \cite{zhao2018beam}. Note that the mechanical approach is often adopted for coarse alignment while the electrical beamforming approach is employed for fine alignment. The mechanical approach adjusts the normal vector of the planar array toward the antenna of the UAV, that is, employing the incident angle to be within a certain range as $[-\theta_a,\theta_a]$ for an example. When the incident angle deviates from the normal vector of the planar array, the beamforming performance declines \cite{zhao2018beam}, and the linearization error of the monopulse signal increases, resulting the performance degradation of the estimation. Therefore, we can expect a stable tracking performance by estimating the fine angle of UAV, after the mechanical adjustment for angle coarsely.

In the proposed scheme, we estimate real-time errors and detect the beam misalignment with the received power in the transmission phase. When the estimated error exceeds a certain threshold, the scheme declares the beam misalignment and executes the mechanical alignment. Note that in this paper, we do not mention the method for an initial estimation and mechanical beam alignment \cite{zhao2018beam}, but we only consider the method for estimating errors and detecting beam misalignment.

The received power can be expressed as a value on the 3-D beam pattern model \cite{huang2010hybrid} as
\begin{equation} \label{eq:Pr}
    P_r=\frac{1}{N^2}\left(\frac{\sin{\left(\frac{N_x}{2}(u-\hat{u})\right)}}{\sin{\left(\frac{1}{2}(u-\hat{u})\right)}}\right)^2\left(\frac{\sin{\left(\frac{N_y}{2}(v-\hat{v})\right)}}{\sin{\left(\frac{1}{2}(v-\hat{v})\right)}}\right)^2.
\end{equation}
In this section, we ignore the subscript of time index $k$ for notational simplicity. By exploiting the cosine pattern approximation of the main lobe of the beam pattern model \cite{deng2018novel} as $\frac{\sin^2{\left(\frac{N}{2}x\right)}}{N^2\sin^2{\left(\frac{x}{2}\right)}}\simeq \cos^2{\left(\frac{N}{4}x\right)}$,
(\ref{eq:Pr}) can be written as
\begin{equation} \label{eq:bp_approx}
\begin{aligned}
    P_r \simeq \cos&{\left(\frac{N_x}{4}(u-\hat{u})\right)}^2 \cos{\left(\frac{N_y}{4}(v-\hat{v})\right)}^2, \\ &u-\frac{2\pi}{N_x}<\hat{u}<u+\frac{2\pi}{N_x}, \\ &v-\frac{2\pi}{N_y}<\hat{v}<v+\frac{2\pi}{N_y},
\end{aligned}
\end{equation}
which is validated in the main lobe. The approximation error can be incurred, but the misalignment detection function is not susceptible to relatively slight errors.
Here, we can represent the estimation error as $\boldsymbol{\xi}=\mathbf{x}-\hat{\mathbf{x}}$. Arranging (\ref{eq:bp_approx}) with $\boldsymbol{\xi}$, we can obtain
\begin{equation} \label{eq:Pr1}
    P_r \simeq \cos{\left(\frac{N_x}{4}\boldsymbol{\xi}_1\right)}^2 \cos{\left(\frac{N_y}{4}\boldsymbol{\xi}_2\right)}^2, |\boldsymbol{\xi}_1|<\frac{2\pi}{N_x}, |\boldsymbol{\xi}_2|<\frac{2\pi}{N_y},
\end{equation}
where $\boldsymbol{\xi}_i$ denotes the $i$th element of the corresponding vector, and $\frac{2\pi}{N_x}$, and $\frac{2\pi}{N_y}$ are the corresponding null-to-null beamwidth of the main lobe, respectively. To find the error norm for the square array, we reformulate (\ref{eq:Pr1}) as 
\begin{equation} \label{eq:Pr2}
    P_r \simeq\cos{\left(\frac{N_x}{4}\lVert\boldsymbol{\xi}\rVert\right)}^4, \lVert\boldsymbol{\xi}\rVert<\frac{2\pi}{N_x},
\end{equation}
where the three-dimensional beam pattern model has the same magnitude at a location with the same radius from the center, $(0,0)$, then, we can re-express (\ref{eq:Pr1}) as (\ref{eq:Pr2}) using the norm of error, $\lVert\boldsymbol{\xi}\rVert$. Under this approximation, the problem of finding the estimation error can be written as
\begin{equation} \label{eq:prob}
    \hat{\lVert \boldsymbol{\xi} \rVert}=\underset{\lVert \boldsymbol{\xi} \rVert\in E}{\arg \min} \left\vert P_r-f(\lVert \boldsymbol{\xi} \rVert)^2 \right\vert,
\end{equation}
where $f(\lVert \boldsymbol{\xi} \rVert)$ is defined as $\cos{\left(\frac{N_x}{4}\lVert \boldsymbol{\xi} \rVert\right)}^2$. We can solve (\ref{eq:prob}) by an exhausted search over the feasible range of $\lVert \boldsymbol{\xi} \rVert$. Note that the feasible region, $E$ of $\lVert \boldsymbol{\xi} \rVert$ is confined on the line of $E=\{y=x, 0\le x \le \gamma\}$, in which $\gamma$ is below the value of $\frac{2\pi}{N_x}$, which is a null point. The Euclidean distance from $\mathbf{x}$ to $\hat{\mathbf{x}}$ is equal to $\lVert \boldsymbol{\xi} \rVert$, and we search the nearest point on grid line, $(y,y), 0\le y \le \gamma$, with the corresponding radius of $\lVert \boldsymbol{\xi} \rVert$.  With the grid search in the range of $E=[0:\Delta:\gamma]$, we can estimate the estimation error in real-time and detect the beam misalignment when the estimated error is over the threshold such as $3$dB beamwidth, $\frac{0.89\pi}{N_x}$. Moreover, in a rectangular array, we can obtain the estimated error with the same method by extending the corresponding feasible region as $(x,y)$ of $0\le x,y \le \gamma$.

\section{Mean Squared Error Bound Analysis} \label{body_analy}
In this section, we analyze the stability of the proposed scheme by showing that the MSE of the proposed scheme is bounded. One of the methods for analyzing the stability of the system is to calculate the estimated error or boundary of the Lyapunov function \cite{reif1999stochastic}, and we herein use the estimated error dynamics to confirm stability. 

Let us define the estimation error of the spatial direction of UAV's transmitted signal as
\begin{equation} \label{eq:err_dy1}
    \boldsymbol{\xi}_k=\mathbf{x}_k-\hat{\mathbf{x}}_k,
\end{equation}
in here, and then by substituting (\ref{eq:G}) and (\ref{eq:update}) into (\ref{eq:err_dy1}), then (\ref{eq:err_dy1}) can be represented as
\begin{equation} \label{eq:err_dy2}
    \boldsymbol{\xi}_k=\left(\mathbf{I}-\mathbf{K}_k\mathbf{G}\right)\left(\mathbf{x}_k-\hat{\mathbf{x}}_k^-\right)-\mathbf{K}_k\mathbf{n}_{m,k}-\mathbf{K}_k\chi\left(\mathbf{x}_k-\hat{\mathbf{x}}_k^-\right),
\end{equation}
where $\chi\left(\mathbf{x}_k-\hat{\mathbf{x}}_k^-\right)$ denotes remainder terms of Taylor expansion in (\ref{eq:G}), which is negligible in high SNR channel.
Arranging (\ref{eq:err_dy2}) with (\ref{eq:state}) and (\ref{eq:pred}), (\ref{eq:err_dy2}) can be written in recursive form as
\begin{equation}
    \begin{aligned}
     \boldsymbol{\xi}_k& \simeq \left(\mathbf{I}-\mathbf{K}_k\mathbf{G}\right)\left(\mathbf{F}\boldsymbol{\xi}_{k-1}+\mathbf{w}_{k-1}\right)-\mathbf{K}_k\mathbf{n}_{m,k}.
    \end{aligned}
\end{equation}
Then, we can derive the approximated MSE as
\begin{equation} \label{eq:mse1}
    \begin{aligned}
    \mathbb{E}&\left[\lVert\boldsymbol{\xi}_k\rVert^2\right]=\mathbb{E}\left[\boldsymbol{\xi}_k^T\boldsymbol{\xi}_k\right], \\
    &\simeq\mathbb{E}\left[\boldsymbol{\xi}_{k-1}^T\mathbf{A}_k^T\mathbf{A}_k\boldsymbol{\xi}_{k-1}\!\!+\mathbf{w}_{k-1}^T\mathbf{B}_k^T\mathbf{B}_k\mathbf{w}_{k-1}\!\!+\mathbf{n}_{m,k}^T\mathbf{K}_k^T\mathbf{K}_k\mathbf{n}_{m,k}\right],
    \end{aligned}
\end{equation}
where $\mathbf{A}_k$ denotes $\left(\mathbf{I}-\mathbf{K}_k\mathbf{G}\right)\mathbf{F}$, and $\mathbf{B}_k$ is $\left(\mathbf{I}-\mathbf{K}_k\mathbf{G}\right)$.  Using linearity of the $\mathrm{trace}$ operator, (\ref{eq:mse1}) can be rearranged as
\begin{equation} \label{eq:mse}
    \begin{aligned}
    \mathbb{E}\left[\lVert\boldsymbol{\xi}_k\rVert^2\right]
    \simeq&\mathrm{Tr}\left(\mathbb{E}\left(\mathbf{A}_k\mathbf{P}_{k-1}\mathbf{A}_k^T\right)\right) +\mathrm{Tr}\left(\mathbb{E}\left(\mathbf{B}_k\mathbf{Q}_{p,k-1}\mathbf{B}_k^T\right)\right) \\
    &+\mathrm{Tr}\left(\mathbb{E}\left(\mathbf{K}_k\mathbf{Q}_{n,k}\mathbf{K}_k^T\right)\right).
    \end{aligned}
\end{equation}
Note that we herein only consider the stable situation to evaluate the accuracy of estimation, thus we assume that process and measurement noises are constant as $\mathbf{Q}_p=\mathbf{Q}_{p,k}$ and $\mathbf{Q}_n=\mathbf{Q}_{n,k}$, for any $k$, over time while UAV moves. Moreover, the process noise variances are known and determined as $\sigma^2_u$ and $\sigma^2_v$, but the measurement noise variance $\sigma_n^2$ of $\mathbf{n}_{m,k}$ in (\ref{eq:err_dy2}) is unknown \cite{rhudy2012relaxation}, which is a variance of the monopulse signal. Therefore, we approximate the measurement noise variance, $\mathbf{Q}_n$ based on previously measured signals.
Consequently, we can show that (\ref{eq:mse}) is upper bounded as
\begin{equation} \label{eq:mse_ub}
    \begin{aligned}
    \mathbb{E}\left[\lVert\boldsymbol{\xi}_k\rVert^2\right]<&\mathrm{Tr}\left(\mathbb{E}\left(\mathbf{A}_k\mathbf{P}_{k-1}\mathbf{A}_k^T\right)\right) +\mathrm{Tr}\left(\mathbb{E}\left(\mathbf{B}_k\mathbf{Q}_{p}\mathbf{B}_k^T\right)\right) \\
    &+\mathrm{Tr}\left(\mathbb{E}\left(\mathbf{K}_k\mathbf{Q}_{n}'\mathbf{K}_k^T\right)\right),
    \end{aligned}
\end{equation}
where $\mathbf{Q}_{n}'$ is defined as relaxed assumption of $\mathbf{Q}_{n}$ as $\mathbf{Q}_{n}'>\mathbf{Q}_{n}$. Note that the gap between the upper bound of MSE and $\mathbb{E}\left[\lVert\boldsymbol{\xi}_k\rVert^2\right]$ can be approximated to the amount of $\mathrm{Tr}\left(\mathbb{E}\left(\mathbf{K}_k\left(\mathbf{Q}_{n}'-\mathbf{Q}_{n}\right)\mathbf{K}_k^T\right)\right)$.

\section{Simulation Results} \label{simulation}
In this section, we verify from the simulations that the proposed scheme is robust in terms of MSE for UAV beam tracking compared to the codebook based beamforming scheme. We consider a GS with an $N_x\times N_y$ planar array and height $h$, which is eight times the flying radius $R_o$, as an example. Thus, the fixed elevation angle is $0.1244$ radian by assuming height is $8$ times of the radius, which is an stable alignment situation. The azimuth angle is randomly distributed from $[-30, 30]$ degrees considering the mechanical alignment. As a baseline, we consider the codebook-based beamforming and auxiliary beam pair schemes where the number of beamforming vectors across 2-D beam space is set to $K^2$, which is the same as $N_x\times N_y$ \cite{kim2019robust,zhu2017auxiliary}. The process noise matrix $\mathbf{Q}_p$ is assumed as
\begin{equation}
    \mathbf{Q}_p = \begin{bmatrix}\sigma_u^2& 0\\ 0 & \sigma_v^2\end{bmatrix},
\end{equation}
and the measurement noise matrix $\mathbf{Q}_n$ is assumed as $\sigma_n^2\mathbf{I}_m$, where $m$ is the dimension of measurement signal. We present the complexity comparison for three schemes briefly in Table \ref{tb:comp}. The computational complexity for KF can be derived from the dimension of the measurement. The pilot overhead required to build the measurement model of KF can be expressed by the signal slot length, $T_s$. We can see that the proposed scheme has lower complexity than the existing conventional schemes. 
\begin{table}[]
\caption{Comparison of the complexity for the proposed scheme with other schemes.}
\centering
\begin{tabular}{|l|l|l|}
\hline
& Computation & \begin{tabular}[c]{@{}l@{}}Pilot overhead\end{tabular} \\ \hline
\begin{tabular}[c]{@{}l@{}}Codebook based\\ beamforming\end{tabular} &  $\mathcal{O}\left((2K^2)^3\right)$ & $K^2T_s$\\\hline
\begin{tabular}[c]{@{}l@{}}Auxiliary beam\\ pair\end{tabular}  &$\mathcal{O}(2^3)$ & $K^2T_s$ \\\hline
Proposed & $\mathcal{O}(2^3)$ & $T_s$ \\ \hline
\end{tabular}
\label{tb:comp}
\end{table}
\begin{figure}
    \centering
    \subcaptionbox{Beam Tracking with $\mathrm{SNR}=30\mathrm{dB}$ and $\sigma_u, \sigma_v=0.005$}{\includegraphics[width=0.45\textwidth]{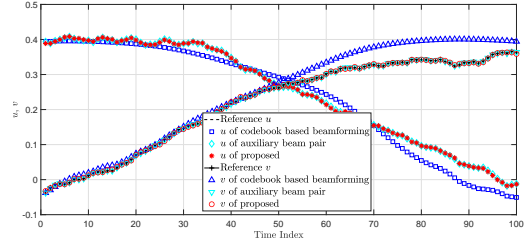}}
    \centering
    \subcaptionbox{Beam Tracking with $\mathrm{SNR}=10\mathrm{dB}$ and $\sigma_u, \sigma_v=0.01$}{\includegraphics[width=0.45\textwidth]{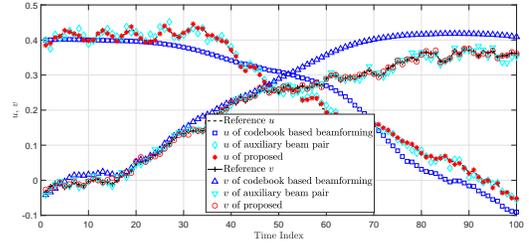}}
\caption{Real-time beam tracking for proposed scheme versus other conventional schemes \cite{zhang2016tracking,kim2019robust} with $u=\pi\cos{\phi}\sin{\theta}$ and $v=\pi\sin{\phi}\sin{\theta}$ based on $8\times 8$ UPA.}
\label{fig:BT}
\end{figure}

Fig. \ref{fig:BT} shows the tracking performance in real-time compared to the conventional beamforming schemes. It is observed that the proposed and auxiliary beam pair schemes can track perturbations of the spatial directions more accurately while the codebook based beamforming is not capable of tracking the abrupt beam changes rapidly from the measurement. We can see from Fig. \ref{fig:BT} that the codebook based beamforming scheme estimates the angle depending on the state transition model without a correct update from measurements. Thus, the tracking performance of the codebook based beamforming scheme gradually degrades as the estimation error accumulates over time since the estimated state can be affected by the previously estimated state. Moreover, the proposed scheme shows more accurate tracking than the auxiliary beam pair scheme in a more dynamic channel as in Fig. \ref{fig:BT} (b).

\begin{figure}[htbp]
    \centering
    \includegraphics[width=0.45\textwidth]{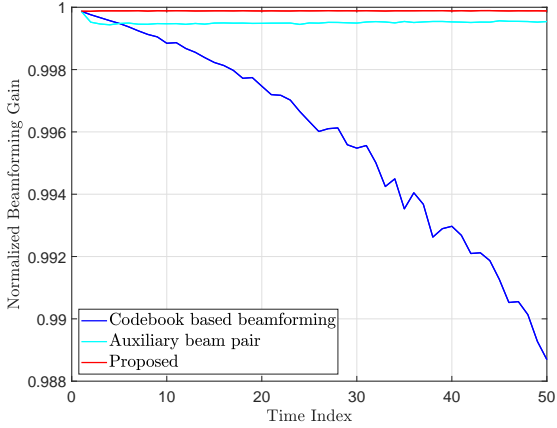}
    \caption{Normalized beamforing gain for proposed scheme versus other conventional beamforming schemes \cite{zhang2016tracking,kim2019robust} with SNR$=10$ [dB] and $\sigma_u, \sigma_v=0.005$ based on $8\times 8$ UPA.}
    \label{fig:bfgain}
\end{figure}

In Fig. \ref{fig:bfgain}, we show the trends of the 3D normalized beamforming gain over time for the proposed scheme and other beam tracking schemes to show a beam alignment accuracy in the data transmission phase. We observe from Fig. \ref{fig:bfgain} that the proposed and auxiliary beam pair schemes achieve a consistent gain based on the robust estimation, while the performance of codebook based beamforming scheme degrades with the increase of time index. 

\begin{figure}
    \centering
    \subcaptionbox{Beam Tracking Detection}{\includegraphics[width=0.45\textwidth]{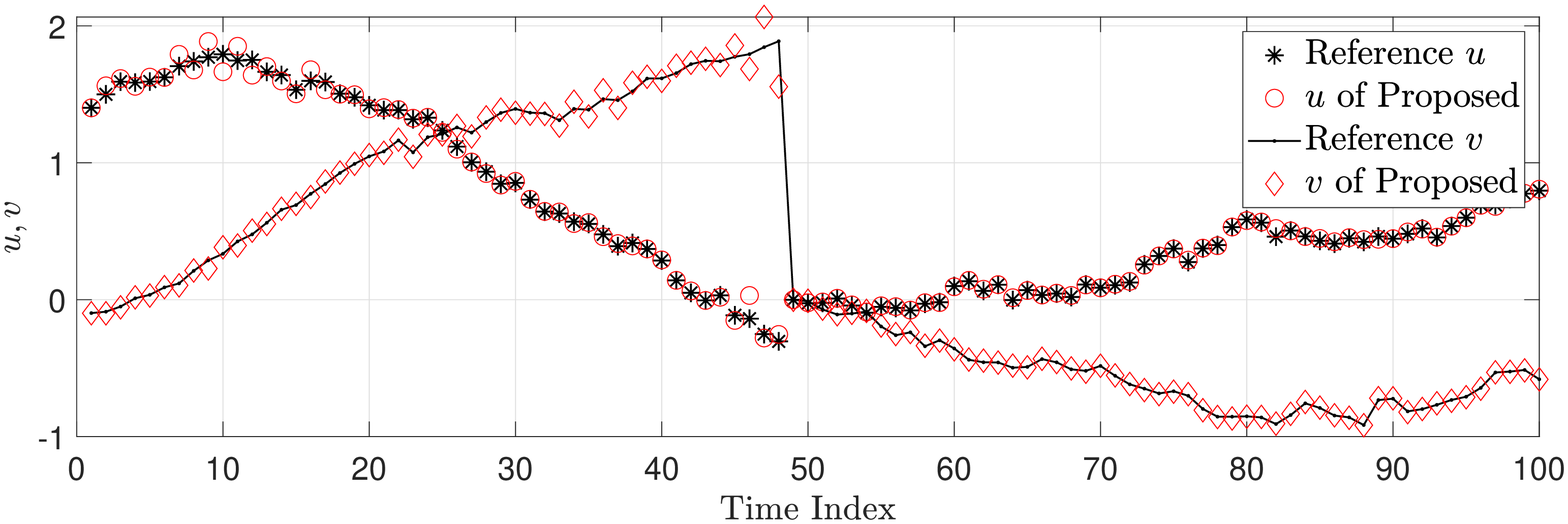}}
    \centering
    \subcaptionbox{Error Estimation}{\includegraphics[width=0.45\textwidth]{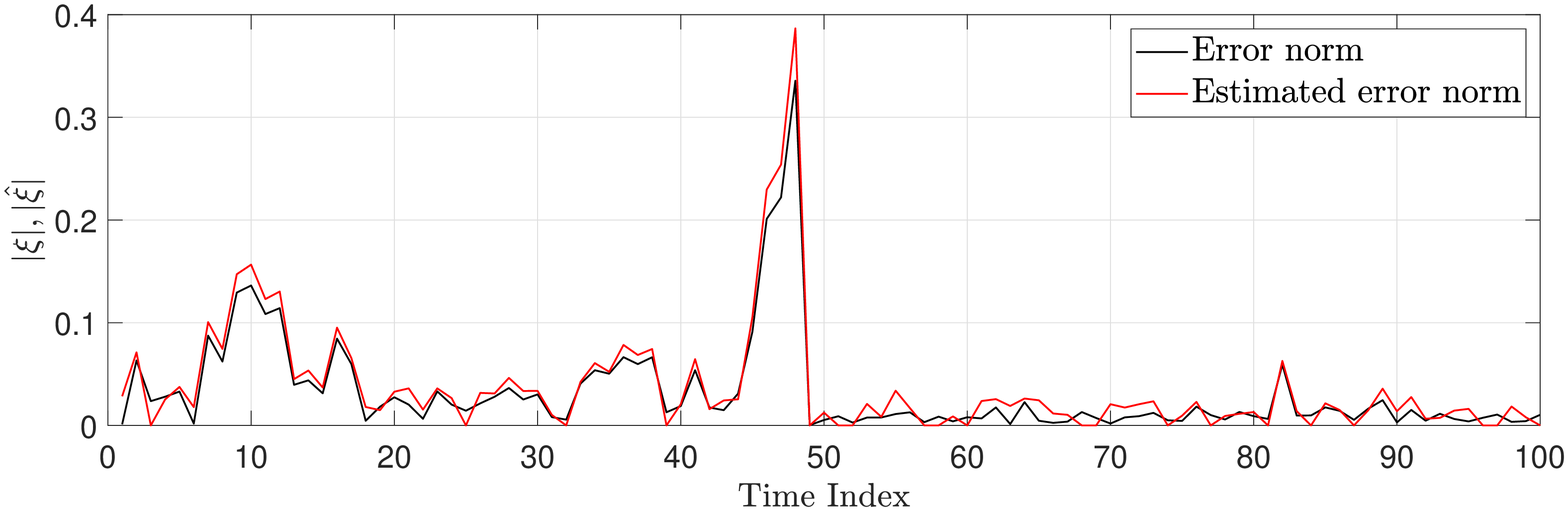}}
\caption{Beam misalignment detection for proposed scheme with $\sigma_u, \sigma_v=0.05$, SNR as $30$ [dB], and $8\times 8$ UPA.}
\label{fig:er_dt}
\end{figure}
In Fig. \ref{fig:er_dt}, we present the tendency of the estimation error. We set the value of SNR as $30$dB, and the process noise standard deviation as $\sigma_u, \sigma_v=0.05$ for an example. The assumed variation in here is large compared to the assumptions considered in \cite{zhang2016tracking,va2016beam,liu2018channel}, thus it can detect the fluctuations in general navigations. Moreover, we also set the height as the two times of the navigation radius as $h=2\times R_o$, which is a misalignment situation with the UAV deviating from the normal vector of the planar array. 
At each frame, the system checks the beam alignment with beamforming power. We can observe from Fig. \ref{fig:er_dt} that the estimation performance degrades from the $45$th time index. In this example, at the $48$th time index, the estimated error exceeds the $3$dB beamwidth, then the scheme declares the misalignment. As a result, the scheme adjusts the beam with the mechanical alignment and can track the UAV with stable performance. After the mechanical alignment, we can see that the spatial angles, $u$ and $v$ become small values around $0$.

\begin{figure}[htbp]
    \centering
    \includegraphics[width=0.45\textwidth]{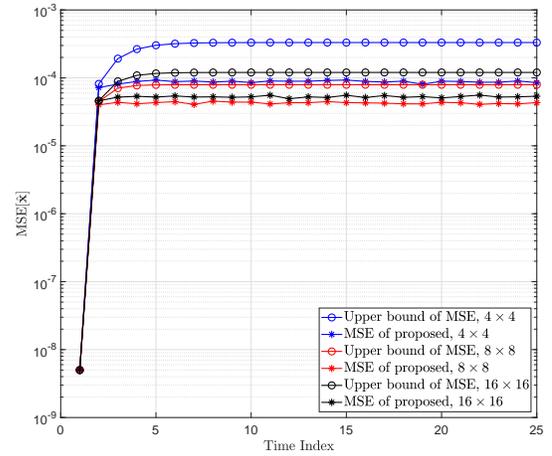}
    \caption{Mean squared error for proposed scheme and mean squared error bound for $N_x=N_y=4$, $N_x=N_y=8$, and $N_x=N_y=16$ with SNR $=10$ [dB] and $\sigma_u, \sigma_v=0.005$.}
    \label{fig:mseb}
\end{figure}

\begin{figure}[htbp]
    \centering
    \includegraphics[width=0.45\textwidth]{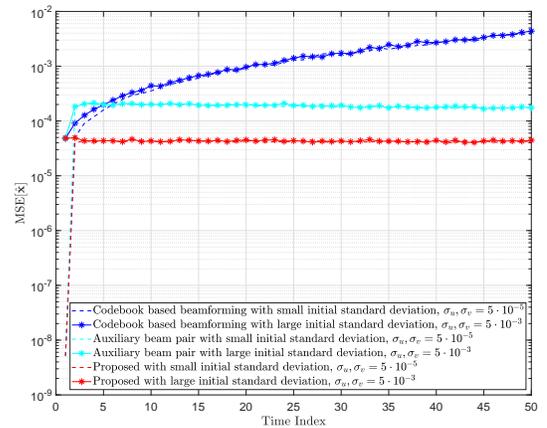}
    \caption{Mean squared error for proposed scheme and other conventional beamforming schemes \cite{zhang2016tracking,kim2019robust} for relatively small/large initial estimation errors, $\sigma_u,\sigma_v=0.00005$ and $\sigma_u,\sigma_v=0.005$ based on $8\times 8$ UPA and SNR $=10$ [dB].}
    \label{fig:mse_initial}
\end{figure}

\begin{figure}[htbp]
    \centering
    \includegraphics[width=0.45\textwidth]{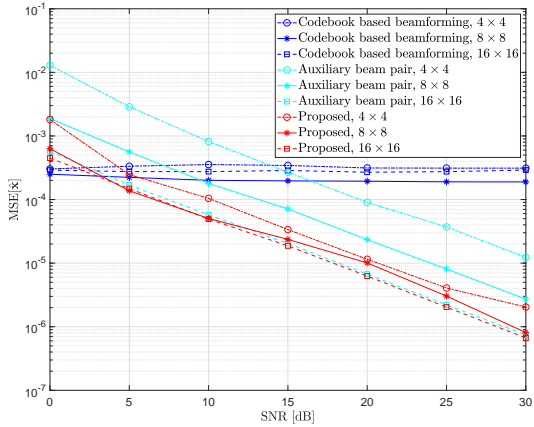}
    \caption{Mean squared error for proposed scheme versus other conventional beamforming schemes \cite{zhang2016tracking,kim2019robust} under various SNR levels and the number of antennas as $N_x=N_y=4$, $N_x=N_y=8$, and $N_x=N_y=16$.}
    \label{fig:mse_snr}
\end{figure}

The Fig. \ref{fig:mseb} shows MSE uppper bound and MSE for the various number of antennas such as $N_x=N_y=4$, $N_x=N_y=8$, and $N_x=N_y=16$. We consider the case that the height is $h=8\times R_o$, the process standard deviation is $\sigma_u,\sigma_v=0.005$, and the initial standard deviation is $\sigma_u,\sigma_v=0.00005$, which are suitable considering the angular change. To deal with a numerical MSE bound, we establish the relaxed measurement noise matrix $\mathbf{Q_n}'$ bigger enough than $\mathbf{Q_n}$ to cancel out the errors from remaining terms of Taylor expansion. The gap between the upper bound of MSE and MSE results embraces the amount from measurement noise variance difference, the linearization loss of Jacobian matrix, and the remainders of Taylor expansion in the measurement model.  
It should be noted that the exact variance of the monopulse signal may not be available, thus we approximate the value of $\sigma_n^2$ as $5\cdot 10^{-6}$ and the relaxed assumption $\sigma_{nb}^{2}$ as $3\cdot10^{-5}$ for $N_x=N_y=8$ case for an example. In addition, we can observe that the estimation accuracy for $N_x=N_y=8$ case is better than that of $N_x=N_y=16$ case, resulting from an angular ambiguity problem \cite{huang2010hybrid}. The antenna element phase can be over $\pi/2$ in the large array, resulting in the performance degradation of the proposed scheme. From Fig. \ref{fig:mseb}, we can verify that the proposed scheme is stable over time with the obtained MSE upper bound.

In Fig. \ref{fig:mse_initial}, we present MSE performance versus the initial estimation errors. Regarding the proposed and auxiliary beam pair schemes, the tracking accuracy of the first and the second time indexes depend on the initial estimation error, but after that, the performance trend is consistent regardless of the initial estimation error. It can be shown that codebook based beamforming scheme is more susceptible to the initial error up to around the $10$th time index. 
Moreover, it is clearly seen that the proposed scheme outperforms the conventional schemes even with lower complexity.

From Fig. \ref{fig:mse_snr}, we also present MSE performance according to the SNR region, where the measurement noise is set as the value related to the noise variance of the received signals. For various SNR regions, it is observed that the auxiliary beam pair scheme shows comparable performance to the proposed scheme in the $16\times 16$ case. However, the proposed scheme is superior with even less complexity to the conventional schemes. We can see that, to improve MSE performance for the proposed one from $10^{-4}$ to $10^{-5}$, it requires about $12$dB in the $8\times 8$ case. Furthermore, we point out that Fig. \ref{fig:mse_snr} shows the limitations of using a beamformed signal as a measurement model for a KF based scheme without performance improvement, with respect to SNR. 
Therefore, we can conclude that the proposed scheme is appropriate for KF based tracking system.

\section{Conclusions} \label{conclusion}
In this study, we proposed a robust UAV tracking scheme as an EKF based beam tracking algorithm for UAV-based air-network. First, we presented the limitations of conventional beam tracking studies and designed a more plausible framework to solve this problem. Simulation results showed that the proposed scheme accurately tracks the perturbation of beam direction of UAV using the monopulse signal with low complexity and estimates the error to detect beam misalignment in real-time. Moreover, we derived the numerical MSE upper bound to analyze the performance and then demonstrated that the proposed scheme is stable over time while UAV keeps moving and outperforms conventional one.

\section*{Acknowledgement}
The authors would like to thank Dr. Hyun-Ho Lee for his constructive comments that greatly improved the quality of this work. This work was supported by Agency for Defense Development(ADD)(UD190008ED).

\bibliography{IEEEabrv,biblist}
\bibliographystyle{IEEEtran}

\end{document}